# Toward possibility of high-temperature bipolaronic superconductivity in boron tubular polymorph: Theoretical aspects of transition into anti-adiabatic state

P. Baňacký[1*], J. Noga[2,3], V. Szöcs[1]

[1]Chemical Physics Division, Institute of Chemistry, Faculty of Natural Sciences, Comenius University, Mlynska dolina CH2, 84215 Bratislava, Slovakia
[2]Department of Inorganic Chemistry, Faculty of Natural Sciences, Comenius University, Mlynska dolina CH2, 84215 Bratislava, Slovakia
[3]Institute of Inorganic Chemistry, Slovak Academy of Sciences, Dubravska cesta 9, 84536 Bratislava, Slovakia

**Abstract.** Large diameter single-wall boron nanotubes (SWBNT) produced by 2%Mg-mesoporous $Al_2O_3$ catalysis show diamagnetic transition at ~ 40 K and ~ 80 K, which is a serious indication for possible superconductivity (J.Phys.Chem.C113, (2009) 17661). Theoretical study which explains or disproves possibility of superconductivity in boron is so far absent, however. Here we apply first-principles formulation of nonadiabatic theory of electron-vibration interactions in study of band structure of boron nanotubes. The *ab initio* results show that electron-vibration coupling induces in SWBNT with diameter larger than 15 Å transition into anti-adiabatic ground state at distorted-fluxional geometry. Thermodynamic and magnetic properties of anti-adiabatic ground state imply possibility of bipolaronic superconductivity. Calculated critical temperature $T_c$ of large diameter SWBNT is 39 K and inclusion of Mg into a tube increases $T_c$ up to 70-90 K. Presence of Al in SWBNT suppress superconductivity and a tube remains metallic down to 0 K. Superconducting properties could established boron nanotubes superior material for nanotechnology.

**Key Words:** A. Superconductors; A. Nanostructures; C. Ab initio calculations; D. Electronic structure; D. Superconductivity

**PACS number(s):**74.10.+v, 73.22.-f, 74.20.-z, 74.20.Mn

## 1. Introduction

Because of their expected electronic properties due to the presence of multicenter bonds, the boron nanotubes and planar structures may be attractive alternatives to carbon tubular [1] and 2-D structures [2] in rapidly growing area of nanotechnology. Scarce experimental output related to these nanostructure forms of boron might be caused by a complex crystal structure and the fact that the phase diagram for boron has been unclear until recently [3]. Moreover, unlike for 2-D graphene, stable single layer precursor form has not yet been reported. Nonetheless, synthesis of large diameter single-wall boron nanotubes (SWBNT) has been successful [4]. Afterwards, DFT-based studies [5,6] proposed that most stable planar structure for SWBNT formation should be the B-α sheet of hexagonal primitive (hP) lattice with 8 B-atoms/unit cell (Bα8). Band structure calculation [6] of Bα8-SWBNT with diameter larger than ~17Å indicated a metal character. Other studies predict a metallic character for all SWBNT irrespective of their diameter and chirality [7,8], which is in agreement with latest experimental results [9]. Recently synthesized SWBNT of diameter ~40Å, produced by 2%Mg-mesoporous $Al_2O_3$ catalysis, show diamagnetic transition at ~ 40 K and ~ 80 K [10], which is a serious indication for possible superconductivity. If confirmed, this unexpected result should prove SWBNT as a superior material for nanotechnology.

---

[*] Corresponding author, E-mail: banacky@fns.uniba.sk

For theoretical study of superconductivity in SWBNT, two aspects should be emphasized at the beginning. The first one is related to microscopic mechanism of the transition in superconducting state. Experimentally detected temperatures of diamagnetic transition [10] correspond to a class of high temperature superconductors. To consider strong electron correlation mechanism behind the high critical temperature, as generally supposed for high-Tc cuprates, is unreasonable in case of SWBNT, or SWBNT with incorporation of metals like Mg or Al into tubular structure. More realistic in this case is to start with a model based on electron-phonon interactions, like in the case of superconductivity in $MgB_2$ [11]. High critical temperature, $T_c$~ 39K, and confirmed [12] two-gap character of this B-based superconductor, not only restarted superconductivity research in general, but this case has pointed to persisting theoretical problems in understanding of the underlying physics behind the superconducting state transition. Though there is a general consensus on phonon assisted superconductivity in $MgB_2$, theoretical interpretation [13,14] within the strong-coupling theory [15-17] is physically controversial. Electron-phonon (EP) coupling constant of $\lambda$~0.7, which simulates [13,14] $T_c \approx 39K$, is greater than the critical value, i.e. $\lambda \geq 0.5$ for adiabatic or whatever small $\lambda$ for anti-adiabatic state that triggers polaron collapse of band and bipolaron formation [18]. This effect is related to symmetry breaking, i.e. electronic band structure (BS) in equilibrium nuclear arrangement can change its topology due to EP coupling. For $MgB_2$, splitting of degenerated σ-bands in Γ point above the Fermi level (FL) has been reported [19,20] due to the coupling to $E_{2g}$ phonon mode vibrations ($\omega \approx 0.066$ eV). Moreover, already for the zero-point vibration displacements of B atoms, the top of the lower σ-band fluctuates across the FL, which substantially changes the physical character of the system. Whilst the system at equilibrium is adiabatic, $\omega/E_F$~0.15, in EP coupling when top of the fluctuating σ-band crosses FL, the chemical potential $\mu$ ($\equiv$Fermi energy $E_F$) decreases and becomes smaller than ω, i.e. $\omega/E_F >1$. Consequently, the system becomes anti-adiabatic [21-23]. Since Eliashberg's treatment [15] is based on a strict adiabatic assumption, transition into the anti-adiabatic state not only invalidates application of the strong coupling theory (BCS-like theories in general) [15-17], but at the same time it means breakdown of the crucial Born-Oppenheimer approximation (BOA) which has to be valid not only in equilibrium but also over the relevant configuration space including geometry with corresponding vibration displacements. In these circumstances, calculations beyond the BOA are required to allow comparison of theory and experiment in detail [24], such as it enables the anti-adiabatic theory of superconductivity suggested recently [25-27]. The latter theory solves the BOA breakdown by introducing an explicit dependence of the electronic structure on nuclear dynamics. It was shown that a system in anti-adiabatic state can be stabilized by non-adiabatic EP coupling in a distorted geometry and the ground electronic state is geometrically degenerated with fluxional nuclear configuration. Thermodynamic properties of a system in stabilized anti-adiabatic electronic ground state are identical with thermodynamics of a superconducting state. Temperature of adiabatic↔anti-adiabatic state transition corresponds to $T_c$ of superconductors with super-carriers in form of mobile bipolarons, as we have shown for different types of superconductors including $MgB_2$ and high-temperature cuprates [22, 26, 27].

In what follows, we apply the anti-adiabatic theory [25-28] in study of SWBNT. Since it is the topology of the BS close to the FL that is crucial, a semiempirical INDO treatment for the BS calculations [29] using valence Slater-type atomic orbitals (STO) basis as in the aforementioned $MgB_2$ and cuprates studies [22,26,27] is appropriate.

The second important aspect at study electronic properties of boron tubular structures is a method used for a tube formation. A hypothetical most stable Bα8 sheet whose reference hP unit cell comprises 8 B-atoms and 32 STOs will serve as a parent structure to construct Bα8-SWBNT. If a standard chiral method [30] were used for Bα8-SWBNT formation, the tubular unit cell for large diameter (Ø ~ 39 Å) contains at least 46 unit cells of planar hP lattice, i.e. 368 B-atoms. Calculation of BS is then not only very demanding, but it yields an obscure bundle of 1472 bands. What is crucial, however, is the fact that the relation to the parent planar 32 bands - BS is completely lost. More over, there is confusion about the basic character of calculated BS. It has been shown [6] that BS of SWBNT with the same chiral vector calculated by zone folding method is substantially different from the BS obtained by direct first-principle calculation. While the first one is metallic, the first-principle calculation results in semiconducting character. In the present study, instead of commonly used chiral method (mainly for study of carbon nanotubes), we have used the helical method. Employing of the helical symmetry [31] gives rise to preserving this relation and keeps the reference tubular unit cell the same as that of the parent planar structure. Related to a two-dimensional structure characterized by translational vectors **a** and **b**, any helical tube can be created by rolling up a ribbon corresponding to $m_a$ translations of the reference cell along **a** and an infinite translations along **b**. The helix is defined by helical parameters $(m_a, m_b)$ defining a vector $m_a \cdot \mathbf{a} + m_b \cdot \mathbf{b}$ that is rolled up perpendicular to the helical axis. This vector is mapped on a cylinder surface, makes its circumference, and hence determines the diameter of the tube. For $(m_a, 0)$ the reciprocal space is characterized by a pair of $(k_{\Phi r}, k_{tr})$-values, $k_{\Phi r} = r/m_a$ ($r = 0, 1, \ldots, m_a - 1$) and $k_{tr} \in \langle -1/2, 1/2 \rangle$. Now, translations along **a** correspond to rotations by $\Phi_r = 2\pi(r/m_a)$ to which $k_{\Phi r}$ is related. Translations along **b** are mapped as rototranslations and can be treated as true translations in an infinite one-dimensional system, hence giving rise to continuous values for $k_{tr}$ related to reciprocal rototranslations. To model the tube in practical calculations, the rototranslations were terminated after sufficiently large odd number $m_{tr}$ and a cyclic cluster with periodic boundary conditions was constructed that essentially corresponded to the bulk limit.

In this paper we present the results obtained by applications of the helical method at band structure calculations and the anti-adiabatic theory of electronic ground state for study of possible superconductivity of SWBNT. The pure B-tubes and tubes with incorporation of Mg and Al of different diameters and length have been investigated. Obtained results confirm that SWBNT are metallic irrespective of diameter and helical vector parameters. Coupling to respective stretching vibration mode has disclosed fluctuation of band structure topology characteristic for transition into anti-adiabatic state. Based on this transition we have calculated that tubes with the diameter (Ø) larger than ~ 15 Å are superconductors. For SWBNT with a diameter ~39 Å, the calculated $T_c$ is ~ 39K and tube is a multi-gap superconductor. Incorporation of Mg into the tubular structure (Bα8Mg-SWBNT) results in a single-gap superconductor with $T_c$~ 70-93 K, whilst Bα8Al-SWBNT remains metallic down to 0 K.

In order to keep the paper self-contained, a theoretical background is presented at the beginning.

## 2. Theoretical background
### 2.1 Helical symmetry of nanotubes in band structure calculation

The basic idea of accounting for the helical symmetry in nanotubes as suggested in [32] and also worked over more recently in [31,33] and somewhat differently in [34], was closely followed in our implementation. Since some technical details are different and can be written in a simplified manner, we repeat them here in order to clarify the calculated band structures.

In general, any nanotube with a periodic structure can be constructed by rolling up a single sheet of a two-dimensional structure that is finite in one translation direction and infinite in the other one. We shall restrict ourselves to nanotubes created from two-dimensional hexagonal lattice characterized by two equivalent $|a| = |b|$ primitive translational vectors *a* and *b* that contain an angle of $2\pi/3$. Due to our convention, the translations along the direction of *a* will be treated as finite, whereas "infinite" number of translations is assumed along *b*. A nanotube characterized by a general helical vector ($m_a$, $m_b$) is then created from the ribbon that has $m_a$ translations (0, ..., $m_a$-1) along *a* and "infinite" number $m_{tr}$ of translations along *b*. The finite value $m_b$ ($m_b \ll m_{tr}$) of helical vector number along *b* and finite $m_a$ value characterize the first complete thread of the helix. Such a ribbon is rolled up on a cylinder with the diameter

$$d_{NT} = |m_a \boldsymbol{a} + m_b \boldsymbol{b}|/\pi, \qquad (1)$$

which follows from the fact that the helical vector ($m_a$ *a* + $m_b$ *b*) is rolled up perpendicular to the rotation axis and makes the circumference of the cylinder. The irreducible computational unit cell corresponds to that in the two-dimensional structure except for the geometry relaxation due to the curvature. Exactly as in the two-dimensional structure that is infinite in both dimensions, each such unit experiences the same environment. Original translations along *a* and *b* are now transformed to rototranslations ($\hat{\tau}_a, \hat{\tau}_b$) characterized by the pair of operations ($z_a, \varphi_a$) and ($z_b, \varphi_b$) where $z_a, z_b$ are projections of *a* and *b* onto the axis of the nanotube and $\varphi_a, \varphi_b$ is the rotation angle related to this translation. Hence for any point defined in a cylindrical coordinate system ($\rho, \varphi, z$) is,

$$\hat{\tau}_i \equiv (\rho, \phi + \varphi_i, z + z_i) \quad \text{for i = a, b.} \qquad (2)$$

If we relate a pseudo-vectors $t_i$ to these rototranslations, in analogy with the two-dimensional planar lattice we can define reciprocal pseudo-vectors $t^*_i$ such that

$$t^*_i t_j = 2\pi\delta_{ij}. \qquad (3)$$

Let the atomic orbital $\chi_{j_a,j_b}$ be a counterpart of the reference unit cell atomic orbital $\chi_{0,0}$ in the unit cell defined by $\hat{\tau}_a^{j_a}$ and $\hat{\tau}_b^{j_b}$ rototranslations. The structure created by $m_a$ rototranslations $\hat{\tau}_a$ (including 0) of the reference computational cell can be treated as an ideal cyclic cluster with periodic boundary conditions, since, indeed in the nanotube each unit has an equivalent surrounding. Consequently, from $m_a$ atomic orbitals $\chi_{j_a,0}$ ($j_a = 0$, $m_a - 1$) one can create $m_a$ symmetry orbitals:

$$\chi^{\Phi_r} = \frac{1}{\sqrt{m_a}} \sum_{j_a=0}^{m_a-1} e^{i k_{\Phi_r} \cdot R_{j_a}} \chi_{j_a,0} \quad , \qquad (4)$$

where $R_{j_a} = j_a t_a$ and there are $m_a$ allowed discrete values of $k_{\Phi_r} = (r/m_a) t^*_a$ for $r = 0,..,m_a-1$. These symmetry orbitals are propagated due to the $\hat{\tau}_b$ to "infinity" ($m_{tr}=N \gg m_a, m_b$) providing Bloch orbitals:

$$\chi^{(k_{tr},k_{\Phi_r})} = \lim_{N\to\infty} \frac{1}{\sqrt{m_a \cdot N}} \sum_{j_b=-N/2}^{N/2} \sum_{j_a=0}^{m_a-1} e^{i(k_{tr} \cdot R_{j_b} + k_{\Phi_r} \cdot R_{j_a})} \chi_{j_a,0} \qquad (5)$$

where $R_{j_b} = j_b t_b$ and $k$'s are any values from the first Brillouin zone for the one-dimensional system. In the practical implementation into the codes that generate integrals in a Cartesian coordinate system, *one has to take care for the appropriate rotation of the coordinate system for each basis function center to preserve the rotational symmetry.* Expressed explicitly, to rotation and rototranslation operations have to be subjected not only nuclear centers that results in set of nuclear coordinates of involved atoms on surface of tube, but to these operations all basis functions of involved atoms must be subjected as well, in order to ensure correct directional arrangements of p,d,f-AOs on tubular surface.

Since in this paper we only report on nanotubes with helical vectors $(m_a,0)$, $\mathbf{k}_{\Phi_r} = k_{\Phi_r} t^*_a$ is related to the true rotational angle in the $m_a$-fold symmetry, whereas $\mathbf{k}_{tr} = k_{tr} t^*_b$. In this special case, the rotational symmetry can be as well accounted for through the point symmetry operations as in Ref [34].

## 2.2 Influence of nuclear dynamics on electronic structure – problem beyond the Born-Oppenheimer approximation

Treatment of this problem is presented in full extension in [25-28]. In order to make the results of this paper intelligible, we present here the final form of corrections to particular energy terms, which arise as the consequence of non-adiabatic electron-vibration coupling and explicitly accounts for dependence of electronic structure on instantaneous nuclear coordinates (Q) and momenta (P), i.e. the Q,P-operators dependence. Corrections are with respect to corresponding energy terms calculated for system within the crude-adiabatic approximation, i.e. standard solution of electronic structure calculation at frozen-clumped nuclear configuration within the Born-Oppenheimer approximation.

**a/ Correction to zero-particle term, i.e. correction to the ground-state energy**
Correction to the electronic ground-state energy in the $k$-space representation due to interaction of pair of states mediated by the phonon mode $r$ is,

$$\Delta E^0_{(na)} = 2\sum_{\varphi_{Rk}}\sum_{\varphi_{Sk'}} \int_0^{\varepsilon_{k',max}} n_{\varepsilon_{k'}}(1-f_{\varepsilon^0_{k'}})d\varepsilon^0_{k'} \int_{\varepsilon_{k,min}}^{\varepsilon_{k,max}} f_{\varepsilon^0_k}|u^r_{k-k'}|^2 n_{\varepsilon_k} \frac{\hbar\omega_r}{(\varepsilon^0_k - \varepsilon^0_{k'})^2 - (\hbar\omega_r)^2} d\varepsilon^0_k, \varphi_{Rk} \neq \varphi_{Sk'} \tag{6}$$

In general, all bands of 1$^{st}$ BZ of a multi-band system are covered, including intra-band terms, i.e. $\varphi_{Rk}, \varphi_{Rk'}$ ($k \neq k'$), with energy $\varepsilon^0_k < \varepsilon_F$ of occupied and $\varepsilon^0_{k'} > \varepsilon_F$ of unoccupied states with respect to Fermi level (FL). The Fermi-Dirac populations $f_{\varepsilon^0_k}, f_{\varepsilon^0_{k'}}$ introduce into correction (6) temperature dependence. Term $u^r_{k-k'}$ stands for matrix element of EP coupling and $n_{\varepsilon k}$, $n_{\varepsilon k'}$ are DOS of interacting bands at $\varepsilon^0_{k'}$ and $\varepsilon^0_k$. It is evident that for adiabatic systems, such as metals - $\omega/E_F \ll 1$, this correction is positive and negligibly small. Only for systems in the anti-adiabatic state ($\omega/E_F > 1$) the correction is negative and its absolute value depends on the magnitudes of $u^r_{k-k'}$ and $n_{\varepsilon k}$, $n_{\varepsilon k'}$ at displacement for FL crossing. In the moment when analytic critical point (ACP) of band $\varphi$ approaches FL, the system not only undergoes transition into the anti-adiabatic state (chemical potential, μ≡$E_F$, is now μ<ω) but DOS, $n_\varphi(E_F) = (\partial\varepsilon^0_\varphi/\partial k)^{-1}_{E_F}$, of the fluctuating $\varphi$-band is considerably increased at FL. It is a situation invoking possibility of van Hove singularity formation.
It should be emphasized that the correction (6) characterizes the effect of nuclear dynamics (Q,P) on electronic ground state energy and, within the quazi-particle transformation method applied at formulation of nonadiabatic electron-vibration theory [28] it represents zero-particle terms, i.e. terms without operator-contractions - scalar quantity. In language of perturbation theory scheme it could be paraphrase as zero-order $\langle\Phi_0|H^0|\Phi_0\rangle$ contribution.

**b/ Correction to one-particle terms, i.e. correction to orbital energies**
At transition into the anti-adiabatic state, $k$-dependent gap $\Delta_k(T)$,

$$\Delta(T) = \Delta(0)tgh(\Delta(T)/4k_BT) \tag{7}$$

in quasi-continuum of adiabatic one-electron spectrum is opened. The gap opening is related to shift $\Delta\varepsilon_{Pk}$ of the original adiabatic orbital energies $\varepsilon^0_{Pk}$, $\varepsilon_{Pk} = \varepsilon^0_{Pk} + \Delta\varepsilon_{Pk}$, and to the $k$-dependent change of DOS of particular band(s) at Fermi level. Shift of orbital energies in band $\varphi_P(k)$ in quasi-momentum $k$-space of multi-band solids has the form,

$$\Delta\varepsilon(Pk') = \sum_{Rk'_1 > k_F} |u^{k'-k'_1}|^2 (1 - f_{\varepsilon^0_{k'_1}}) \frac{\hbar\omega_{k'-k'_1}}{(\varepsilon^0_{k'} - \varepsilon^0_{k'_1})^2 - (\hbar\omega_{k'-k'_1})^2} - \sum_{Sk < k_F} |u^{k-k'}|^2 f_{\varepsilon^0_k} \frac{\hbar\omega_{k-k'}}{(\varepsilon^0_{k'} - \varepsilon^0_k)^2 - (\hbar\omega_{k-k'})^2} \tag{8a}$$

for $k' > k_F$, and

$$\Delta\varepsilon(Pk) = \sum_{Rk'_1 > k_F} |u^{k-k'_1}|^2 (1 - f_{\varepsilon^0 k'_1}) \frac{\hbar\omega_{k-k'_1}}{(\varepsilon_k^0 - \varepsilon_{k'_1}^0)^2 - (\hbar\omega_{k-k'_1})^2} - \sum_{Sk_1 < k_F} |u^{k-k_1}|^2 f_{\varepsilon^0 k} \frac{\hbar\omega_{k-k_1}}{(\varepsilon_k^0 - \varepsilon_{k_1}^0)^2 - (\hbar\omega_{k-k_1})^2} \quad (8b)$$

for $k \leq k_F$

Replacement of discrete summation by integration, $\sum_k \ldots \to \int n(\varepsilon_k)$, introduces DOS $n(\varepsilon_k)$ into (8a,b). It is of crucial importance in relation to fluctuating band. For corrected DOS $n(\varepsilon_k)$, which is the consequence of shift $\Delta\varepsilon_k$ of orbital energies, the following relation can be derived;

$$n(\varepsilon_k) = \left|1 + (\partial(\Delta\varepsilon_k)/\partial\varepsilon_k^0)\right|^{-1} n^0(\varepsilon_k^0) \quad (9)$$

Term $n^0(\varepsilon_k^0)$ stands for uncorrected DOS of the original adiabatic states of particular band,

$$n^0(\varepsilon_k^0) = \left|(\partial\varepsilon_k^0/\partial k)\right|^{-1} \quad (10)$$

Close to the **k**-point where the original band which interacts with fluctuating band intersects FL, the occupied states near FL are shifted downward - below FL and unoccupied states are shifted upward - above FL. The gap is identified as the energy distance between created peaks in corrected DOS above FL (half-gap) and below FL. The formation of peaks is related to the spectral weight transfer that is observed for superconductors by ARPES or tunneling spectroscopy in cooling below $T_c$.

With respect to $\Delta(0)$, from (7) a simple approximate relation for critical temperature $T_c$ of the adiabatic↔anti-adiabatic state transition follows,

$$T_c = \Delta(0)/4k_B . \quad (11)$$

**c/ Correction to two-particle term, i.e. correction to the electron correlation energy**

This correction in the **k**-space representation has the form,

$$\Delta H''_{ep} = \sum_{R(k)S(k')kk'q\sigma\sigma'(q \neq 0)} |u^q|^2 \frac{\hbar\omega_q((\varepsilon_{k+q}^0 - \varepsilon_k^0)(\varepsilon_{k'+q}^0 - \varepsilon_{k'}^0) - (\hbar\omega_q)^2)}{((\varepsilon_{k+q}^0 - \varepsilon_k^0)^2 - (\hbar\omega_q)^2)((\varepsilon_{k'+q}^0 - \varepsilon_{k'}^0)^2 - (\hbar\omega_q)^2)} N[a^+_{k+q,\sigma} a^+_{k',\sigma'} a_{k'+q,\sigma'} a_{k,\sigma}] \quad (12)$$

For anti-adiabatic system $|\varepsilon_{k'}^0 - \varepsilon_k^0| < \hbar\omega_{k'-k}$, denominators in (12) are positive and negative value of the matrix elements of this two-particle correction is reached for a reduced form if nominators are negative, i.e. if $(\varepsilon_{k+q}^0 - \varepsilon_k^0) = -(\varepsilon_{k'+q}^0 - \varepsilon_{k'}^0)$. Since $q \neq 0$, it can be reached if $q = -k - k'$. At these circumstances for reduced form which maximizes e-e attraction, is derived

$$\Delta H''_{ep}(red)_{na} = -2 \sum_{k > k_F, k' < k_F} |u^{k'-k}|^2 \frac{\hbar\omega_{k'-k}((\varepsilon_{k'}^0 - \varepsilon_k^0)^2 + (\hbar\omega_{k'-k})^2)}{((\varepsilon_{k'}^0 - \varepsilon_k^0)^2 - (\hbar\omega_{k'-k})^2)^2} N[a^+_{k\uparrow} a^+_{-k'\downarrow} a_{-k\downarrow} a_{k\uparrow}] \quad (13)$$

In this expression, summation over bands is not explicitly indicated, but it should be understand implicitly. This correction is due to pairs of electrons with opposite quasi-momentum and antiparallel spins $(k\uparrow,-k\downarrow)$. It should be noticed that it is the contribution of bi-excited configurations $\{\Phi_{(k',-k')\to(k,-k)}\}$ (i.e. two-particle $(k\uparrow,-k\downarrow)$, two-hole $(k'\uparrow,-k'\downarrow)$ excited singlet states) to the electronic ground state that is represented by renormalized Fermi vacuum $\Phi_0$. Expressed explicitly, first nonzero contributions are from the matrix elements of the type $\langle\Phi_0|\Delta H_{ep}^{''}|\Phi_{(k',-k')\to(k,-k)}\rangle^2$, i.e. contributions arise in the second order of perturbation theory. Now, $\{\varepsilon_k\}$ represent particle states that are occupied above Fermi level and $\{\varepsilon_{k'}\}$ are, due to excitations, empty – hole states below Fermi level.

In strong adiabatic regime, correction to electron correlation energy (reduced form) is small but negative and in the limit $\hbar\omega_{k'-k}/|\varepsilon_{k'}^0-\varepsilon_k^0|\to 0$ it approaches zero-value,

$$\Delta H_{ep}^{''}(red)_{sad} = -2\sum_{kk'}|u^{k'-k}|^2\frac{\hbar\omega_{k'-k}}{(\varepsilon_{k'}-\varepsilon_k)^2}N[a_{k\uparrow}^+ a_{-k\downarrow}^+ a_{-k\downarrow}a_{k\uparrow}]\to 0 \qquad (14)$$

The Fröhlich effective Hamiltonian of e-e interactions,

$$H_{eff}^{''}(Fr) = \sum_{kk'q\sigma\sigma'}|u^q|^2\frac{\hbar\omega_q}{(\varepsilon_{k+q}^0-\varepsilon_k^0)^2-(\hbar\omega_q)^2}a_{k+q,\sigma}^+ a_{k',\sigma'}^+ a_{k'+q,\sigma'}a_{k,\sigma} \qquad (15)$$

is the basis for Cooper pair formation. This interaction term is either attractive or repulsive depending on the sign of denominator. For anti-adiabatic conditions $|\varepsilon_{k'}^0-\varepsilon_k^0|<\hbar\omega_{k'-k}$ it represents effective attractive electron-electron (e-e) interactions. It should be reminded that effective attractive e-e interactions are the crucial condition of Cooper's pair formation and the basis of the BCS theory.

The reduced form of the Fröhlich Hamiltonian is,

$$H_{red}^{''}(Fr) = 2\sum_{kk'}|u^{k'-k}|^2\frac{\hbar\omega_{k'-k}}{(\varepsilon_{k'}^0-\varepsilon_k^0)^2-(\hbar\omega_{k'-k})^2}a_{k'\uparrow}^+ a_{-k'\downarrow}^+ a_{-k\downarrow}a_{k\uparrow} \qquad (16)$$

In the limit of extreme anti-adiabaticity $|\varepsilon_P^0-\varepsilon_Q^0|/\hbar\omega_r\to 0$, the form of the Fröhlich two-particle effective Hamiltonian (16) and the correction to electron correlation energy (13) are identical and equal to,

$$\lim(\Delta H_{ep}^{''})_{\Delta\varepsilon/\hbar\omega\to 0} = \lim(H_{eff}^{''}(Fr))_{\Delta\varepsilon/\hbar\omega\to 0} = -\sum_{kk'q\sigma\sigma'}\frac{|u^q|^2}{\hbar\omega_q}a_{k+q,\sigma}^+ a_{k',\sigma'}^+ a_{k'+q,\sigma'}a_{k,\sigma} \qquad (17)$$

It is evident now, that in anti-adiabatic regime when system is already stabilized (6) at distorted geometry (symmetry breaking), the correction to the two-particle term (13, or 16) represents only contribution in second order of perturbation theory to $\Delta E_{(na)}^0$. Expressed explicitly, Cooper pair formation is a consequence of the transition into anti-adiabatic state and represents an increase of electron correlation energy (13) when system is already stabilized (6), due to nonadiabatic EP coupling, in anti-adiabatic ground electronic state at distorted geometry.

**d/ Physical properties of system in anti-adiabatic ground state**
On crude-adiabatic level (clumped-nuclear situation), total ground state electronic energy $E_0^{te}(R_{eq})$ has minimum at $R_{eq}$, i.e.; $\left(dE_0^{te}/dR\right)_{R_{eq}} = 0$. Related to any phonon mode, nuclear displacements in vibration motion increase total electronic energy (potential energy of nuclear motion in particular phonon mode). For an increase of the total electronic energy $\Delta E_d$ due to nuclear displacement $R_d$ holds,

$$\Delta E_d(R_d) = E_0^{te}(R_d) - E_0^{te}(R_{eq}) > 0 \tag{18}$$

In principle, two situations can occur; nuclear displacements related to some phonon mode(s) induce formation of anti-adiabatic state, or system remains in adiabatic state with respect to vibration motion in all phonon modes. The question if anti-adiabatic state can be a stable state, i.e. to be a ground electronic state of system at distorted nuclear configuration $R_d$, depends on the value of the ground state energy correction (6), $\Delta E_{(na)}^0(R_d)$. Since for anti-adiabatic state this correction is negative, $\Delta E_{(na)}^0(R_d) < 0$, then if the inequality $\left|\Delta E_{(na)}^0(R_d)\right| > \Delta E_d(R_d)$ holds, the electronic state of the system is stabilized at distorted geometry $R_d$. The reason of it is significant participation of the nuclear kinetic energy term expressed through contribution of $P$-dependent transformation, which stabilizes fermionic ground state energy in anti-adiabatic state at distorted nuclear configuration $R_d$.
Stabilization (condensation) energy at transition from adiabatic into anti-adiabatic state is,

$$E_{cond}^0 = \Delta E_d(R_d) - \left|\Delta E_{(na)}^0(R_d)\right| \tag{19}$$

**d1/ Character of condensation in anti-adiabatic state**
In stabilized anti-adiabatic state, ground state total electronic energy of solid state system is geometrically degenerated. Distorted nuclear structure, related to the couple of nuclei in the phonon mode $r$ that induces transition into anti-adiabatic state, has fluxional character. There exist an infinite number of different - distorted configurations of involved couple of nuclei in the phonon mode $r$ and all these configurations, due to translation symmetry of the lattice, have the same ground state energy. Position of displaced couple of nuclei is on the perimeter of circles with the centers at $R_{eq}$ (equilibrium on crude-adiabatic level) and with radii equal to $\Delta R = \left|R_{eq} - R_{d,cr}\right|$. The $R_{d,cr}$ is distorted geometry at which ACP approaches FL and system undergoes transition from adiabatic into anti-adiabatic state. Due to the geometric degeneracy of the ground state energy, the involved atoms can circulate over perimeters of the circles without the energy dissipation. The dissipation-less motion of the couple of nuclei implies, however, that EP coupling of this phonon mode and electrons of corresponding band has to be zero in stabilized anti-adiabatic state. The effective EP interactions which cover the nuclear coordinates and momenta ($Q$, $P$)-dependence has the form

$$\Delta H'_{\langle e-p0\rangle}(dg) = \sum_{k,q}\left|u^q\right|^2\left\{\left(\varepsilon_k^0 - \varepsilon_{k-q}^0\right)\Big/\left[\left(\varepsilon_k^0 - \varepsilon_{k-q}^0\right)^2 - (\hbar\omega_q)^2\right]\right\}N[a_k^+ a_k] \qquad (20)$$

For extreme nonadiabatic limit, e.g. anti-adiabatic state, $\hbar\omega_q / \left|\varepsilon_k^0 - \varepsilon_{k-q}^0\right| \to \infty$, follows

$$\Delta H'_{\langle e-p0\rangle}(dg)_{na} \to 0 \qquad (20a)$$

It means that for electrons which satisfy condition of extreme nonadiabaticity with respect to interacting phonon mode $r$, in particular direction of reciprocal lattice where the gap in one-electron spectrum has been opened, the electron (i.e. nonadiabatic polaron)-renormalized phonon interaction energy equals zero. Expressed explicitly, in the presence of external electric potential, dissipation-less motion of relevant valence band electrons (holes) on the lattice scale can be induced at the Fermi level (electric resistance $\rho = 0$), whilst motion of nuclei remains bound to fluxional revolution over distorted, energetically equivalent, configurations. The electrons move in a form of itinerant bipolarons.

The bipolarons arise as polarized inter-site charge density distribution that can move over lattice without dissipation due to geometric degeneracy (fluxional structure) of the anti-adiabatic ground state at distorted nuclear configurations. Formation of polarized inter-site charge density distribution at transition from adiabatic into anti-adiabatic state is reflected by corresponding change of the wave function. For simplicity, let to consider that transition into anti-adiabatic state is driven by coupling to a phonon mode $r$ with stretching vibration of two atoms (e.g. B-B in $E_{2g}$ mode of MgB$_2$). Let $m_1$ and $m_2$ are equilibrium site positions of involved nuclei on crude-adiabatic level and $d_1$ and $d_2$ are nuclear displacements at which crossing into antiadiabatic state occurs. At these circumstances, the original crude-adiabatic wave function $\varphi_k^0(x,0,0)$ is changed in a following way [26],

$$\left|\varphi_P(x,Q,P)\right\rangle = a_P^+(x,Q,P)|0\rangle = \left(\overline{a}_P^+ - \sum_{rR}c_{PR}^r\overline{Q}_r\overline{a}_R^+ - \sum_{rR}\hat{c}_{PR}^r P_r \overline{a}_R^+ + O(\overline{Q}^2,\overline{Q}P,P^2)\right)|0\rangle =$$
$$= \left|\varphi_P(x,0,0)\right\rangle - \sum_{rR}c_{PR}^r\overline{Q}_r\left|\varphi_R(x,0,0)\right\rangle - \sum_{rR}\hat{c}_{PR}^r\overline{P}_r\left|\varphi_R(x,0,0)\right\rangle + \ldots\ldots \qquad (21)$$

At transition into anti-adiabatic state $\left|\varepsilon_S^0(k_c) - \varepsilon_F^0\right|_{R_{eq}\pm Q} \ll \hbar\omega_r$, coefficients $c_{RS}^r$ of $Q$-dependent transformation matrix become negligibly small and absolutely dominant for modulation of crude-adiabatic wave function are in this case coefficients $\hat{c}_{RS}^r$ of $P$-dependent transformation matrix. Then for wave function holds,

$$\varphi_k(x,Q,P) \propto \left(1 + \sum_q u^{|q|}\frac{\hbar\omega_q}{(\hbar\omega_q)^2 - (\varepsilon_k^0 - \varepsilon_{k+q}^0)^2}\left(P_1 e^{iq\cdot[x-(m_1-d_1)]} + P_2 e^{iq\cdot[x-(m_2+d_2)]}\right)\right)\varphi_k^0(x,0,0) \qquad (21a)$$

In (21a), site approximation for momentum has been used, i.e. $P(q) \propto (sign.q)\sum_m P_m e^{iq\cdot m}$.

In anti-adiabatic state, for particular $k$ and proper $q$ values, nonadiabatic prefactors under summation symbol in (21a) can be large. The prefactors, i.e. contribution of $P$-dependent transformation matrix, reflect influence of nuclear kinetic energy on electronic structure. At the dominance of these contributions (anti-adiabatic state), strong increase in localization of charge density appears at distorted site-positions for $x$ equal to $(m_1-d_1)$ and $(m_2+d_2)$. It induces (or increases) electronic polarization with alternating higher and smaller inter-site charge density distribution. This kind of inter-sites polarization persists, and on a lattice scale it has itinerant character at nuclear revolution over perimeters of the fluxional circles with the radius $R_{d,cr}$, until the system remains in the anti-adiabatic state. Due to temperature increase, thermal excitations of valence band electrons to conduction band induce sudden transition from the anti-adiabatic state into adiabatic state at $T = T_c$, i.e. inequality $\left|\Delta E^0_{(na)}(R_d)\right| > \Delta E_d(R_d)$ does not hold. For temperatures $T \geq T_c$ now holds $\left|\Delta E^0_{(na)}(R_d)\right| \leq \Delta E_d(R_d)$ and system becomes stable at equilibrium $R_{eq}$ as it is characteristic for adiabatic structure.

In the adiabatic state, properties of the electrons are in sharp contrast with the properties of electrons in anti-adiabatic state. The electrons in this case, are in a valence band more or less, tightly bound to respective nuclei at adiabatic equilibrium positions and theirs motion in conducting band is restricted by scattering with interacting phonon modes. It corresponds to situation at $T>T_c$.

For extreme adiabatic limit $\hbar\omega_q / \left|\varepsilon^0_k - \varepsilon^0_{k-q}\right| \to 0$, from (20) for electron-phonon interaction energy in this case follows,

$$\Delta H'_{\langle e-p0\rangle}(dg)_{ad} \to \sum_{qk}\left|u^q\right|^2 \frac{1}{\left(\varepsilon^0_k - \varepsilon^0_{k-q}\right)} \tag{22}$$

Expression (22) represents well known energy of standard adiabatic polarons (small, self-trapped) that contributes to the total energy of system.

**d2/ Electronic specific heat and entropy in anti-adiabatic state**

At formation of the anti-adiabatic ground state, electronic energy is decreased and for involved band(s) the gap in one-particle spectrum has been opened (shift of orbital energies). This fact is reflected by change of related thermodynamic properties. In particular, for electronic specific heat

$$C_{V,el}(T) = \frac{d\Delta E^0_{(na)}}{dT} = T\frac{dS}{dT} \tag{23}$$

can be derived,

$$C_{V,el}(T) = -\frac{\bar{n}(\varepsilon_{k_F})}{2}\frac{\Delta(T)}{\Delta(0)}\left(\frac{\Delta(T)}{dT}\right) \tag{23a}$$

For entropy related to formation of anti-adiabatic state holds,

$$S = k_B \ln 2 - \bar{n}(\varepsilon_{k_F}) \frac{(\Delta(T))^2}{2T\Delta(0)} + 2k_B \bar{n}(\varepsilon_{k_F}) \ln\left( \cosh \frac{\Delta(T)}{4k_B T} \right) \qquad (24)$$

For temperature derivative of entropy (24) follows,

$$\frac{dS}{dT} = -\frac{\bar{n}(\varepsilon_{k_F})}{2T} \frac{\Delta(T)}{\Delta(0)} \frac{d\Delta(T)}{dT} \qquad (24a)$$

From the (23a), in the limit T→ 0 K, the exponential behavior characteristic for superconductors can be derived,

$$\lim(C_{V,el}(T))_{T \to 0} \approx \exp\left( -\frac{\Delta(0)}{2k_B T} \right) \qquad (25)$$

The density of states at Fermi level $\bar{n}(\varepsilon_{k_F})$ in above equations represents mean value of corrected density of states close to the *k*-point where the peak in density of states has been formed. From practical reasons it can be approximated by the mean value of density of states of the fluctuating band in ACP at the moment when it approaches Fermi level at distance of ($\pm \omega/2$) and anti-adiabatic state is established. In anti-adiabatic state, density of states at Fermi level is considerably increased since density of states of fluctuating band at ACP is usually high (possibility of van Hove singularity formation at Fermi level).

**d3/ Magnetic properties of system in the anti-adiabatic state – critical magnetic field**
System in superconducting state can exhibits absolute diamagnetism and Meissner effect only if inside the system $B = 0$. In this case, there has to exists some critical value of external magnetic field $H_c$ which destroy superconducting state and induces transition of the system into normal state (characteristic by finite-nonzero value of electric resistance $\rho \neq 0$ at finite-nonzero density of electric current *j* ), like it occurs in case of temperature increase above $T_c$. It also means that at critical temperature and above it, $T \geq T_c$, critical magnetic field has to be zero, $H_c(T \geq T_c) = 0$.
It can be shown that anti-adiabatic state exhibits this property.
From thermodynamics for critical magnetic field in this case follows,

$$\frac{H_c^2}{8\pi} = F_{(ad)} - F_{(na)} \qquad (26)$$

In (26), $F_{(ad)}$ and $F_{(na)}$ stand for free energies of the system in adiabatic and nonadiabatic (anti-adiabatic) state. For the change in free energy at the transition holds,

$$F_{(ad)} - F_{(na)} = \Delta E_{(na)}^0 - TS \qquad (26a)$$

From (23, 23a) follows,

$$\Delta E^0_{(na)}(T) = \Delta E^0_{(na)}(0) + \int_0^T C_{V(el)}(T)dT = \Delta E^0_{(na)}(0) - \frac{\bar{n}(\varepsilon_{k_F})}{4\Delta(0)}\left((\Delta(T))^2 - (\Delta(0))^2\right) =$$
$$= -\frac{\bar{n}(\varepsilon_{k_F})}{4}\Delta(0) - \frac{\bar{n}(\varepsilon_{k_F})}{4\Delta(0)}\left((\Delta(T))^2 - (\Delta(0))^2\right) \qquad (27)$$

After substitution of (27,24) into (26a) and algebraic rearrangements, for critical magnetic field at finite temperature T follows,

$$\frac{(H_c(T))^2}{8\pi} = -\frac{\bar{n}(\varepsilon_{k_F})}{4}\Delta(0) - \frac{\bar{n}(\varepsilon_{k_F})}{4\Delta(0)}\left((\Delta(T))^2 - (\Delta(0))^2\right) + \frac{\bar{n}(\varepsilon_{k_F})}{2}\frac{(\Delta(T))^2}{\Delta(0)} -$$
$$- 2k_B\bar{n}(\varepsilon_{k_F})T\ln\left(\frac{\Delta(0)}{\left((\Delta(0))^2 - (\Delta(T))^2\right)^{\frac{1}{2}}}\right) \qquad (28)$$

At temperature 0 K, for critical magnetic field results,

$$(H_c(0))^2 = 2\pi\Delta(0)\bar{n}(\varepsilon_{k_F}) \qquad (28a)$$

The relation between critical magnetic fields at finite and zero temperature follows from (28,28a),

$$\left(\frac{H_c(T)}{H_c(0)}\right)^2 = \left[\left(\frac{\Delta(T)}{\Delta(0)}\right)^2 - \frac{8k_BT}{\Delta(0)}\ln\left(\frac{\Delta(0)}{\left((\Delta(0))^2 - (\Delta(T))^2\right)^{\frac{1}{2}}}\right)\right] \qquad (28b)$$

Due to thermal excitations, the anti-adiabatic state is suddenly changed into adiabatic state at critical temperature $T_c$. Since there is no gap in one-electron spectrum in metal-like adiabatic state, $\Delta(T \geq T_c) = 0$, then zero value of critical magnetic field follows directly from (28,28b), i.e. $H_c(T \geq T_c) = 0$

Derived equations show that system in the anti-adiabatic state, beside zero value of electric resistance $\rho = 0$ (dissipation-less motion of bipolarons), has also specific property that is necessary for occurrence of the Meissner effect.

### 3. Results
### 3.1 Parent structures

A hypothetical most stable Bα8 sheet whose reference unit cell comprises 8 B-atoms and 32 STOs serves as a parent structure to construct Bα8-SWBNT. The crystallographic lattice is hexagonal primitive (hP) and optimized lattice parameters for infinite 2D-sheet are a=b=5.2900 Å. The fractional coordinates of B atoms are; B1(1/3,0,0), B2(2/3,0,0), B3(0,1/3,0), B4(0,2/3,0), B5(1/3,1/3,0), B6(1/3,2/3,0), B7(2/3,1/3,0), B8(2/3,2/3,0). The bond length of B-B atoms is 1.7633 Å.

Calculated band structure (32 bands) corresponding to this structure is displayed in Fig.1a. System is metallic in both high-symmetry directions with σ-like bands degeneracy above FL in Γ point. In the Γ-M direction, the $p_z$-like band and four σ bands intersect FL.

The parent structure for modelling a tube with incorporation of metal atom (Met) is constructed by placing Met over the empty hexagon in Bα8 unit cell with fractional coordinates, Met(0,0,1/2). Optimized lattice parameters for Met=Mg are, a=b=5.4138 Å and c=2.7678 Å. Calculated band structure of Bα8Mg layer is also metallic - Fig.1b. Now, the top of σ-like bands is shifted below the FL and $p_z$-like band intersects FL in Γ-M direction with the top in M point above the FL.

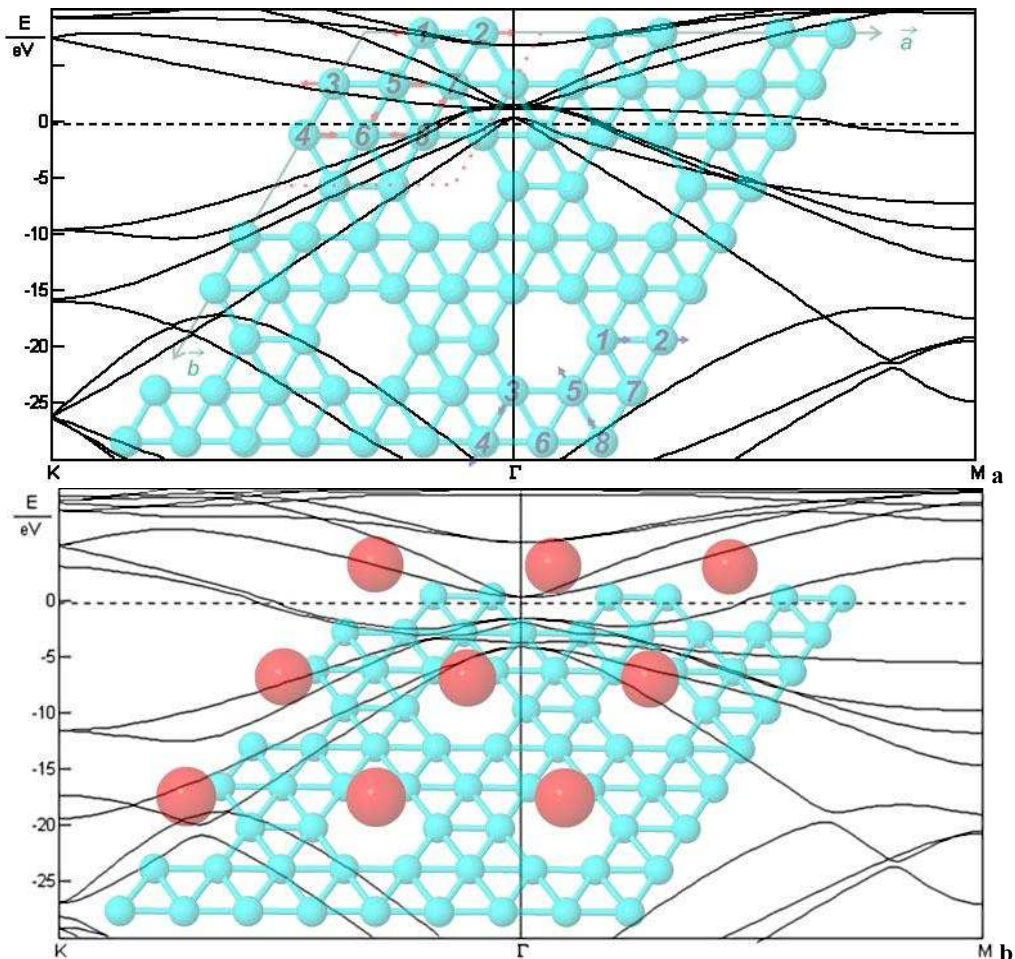

**Figure1. (colour online)** Band structures for Bα8 (a) and Bα8Mg (b) sheets in hexagonal primitive lattice (dotted line is Fermi level). Structural motif is in background with indicated unit cell and B and Mg-atoms. Coordinates of high symmetry points are: Γ=(0,0,0), K=(-2/3,1/3,0), M=(0,1/2,0). The stretching and bending vibrations are indicated (a) by arrows in the upper (red-stretching) and bottom (violet-bending) corner of the motif.

To our knowledge, the phonon spectra of boron tubular structure have not been published so far. In order to be able to study the effects of electron-vibration coupling on electronic structure and to estimate adiabatic ratio $\omega/E_F$ (vide ultra) we have calculated normal modes frequencies for the parent Bα8 sheet and for the basic ring of the tube using DFT-based ADF code [35]. The IR spectra of Bα8 slab have been calculated within the BAND part of the ADF code with GGA-revPBe XC potential. For the B184-ring IR spectra, the ADF part of the code was chosen (with GGA-revPBe XC potential). Calculated IR spectra and table form which illustrates peak positions and corresponding transition dipole strength (available only for ADF results for B184 ring) are presented in Fig.2.

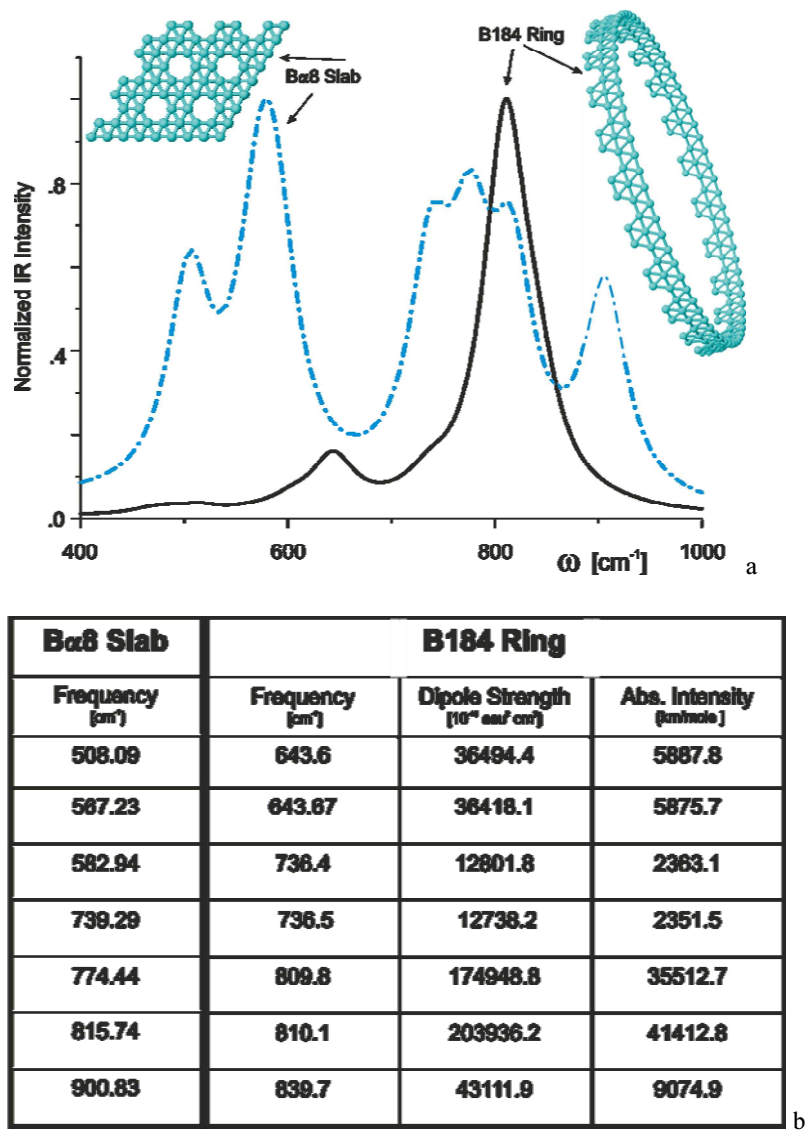

| Bα8 Slab | B184 Ring | | |
|---|---|---|---|
| Frequency [cm⁻¹] | Frequency [cm⁻¹] | Dipole Strength [10⁻⁴ esu² cm²] | Abs. Intensity [km/mole] |
| 508.09 | 643.6 | 36494.4 | 5887.8 |
| 567.23 | 643.67 | 36418.1 | 5875.7 |
| 582.94 | 736.4 | 12801.8 | 2363.1 |
| 739.29 | 736.5 | 12738.2 | 2351.5 |
| 774.44 | 809.8 | 174948.8 | 35512.7 |
| 815.74 | 810.1 | 203936.2 | 41412.8 |
| 900.83 | 839.7 | 43111.9 | 9074.9 |

**Figure 2. (colour online)** Calculated normalized IR intensities of Bα8-slab (dot-dashed blue line) and B184-ring (black line). The figure is complemented with respective structural motifs and table of non-zero normal modes.

For purpose of the present study, relatively rough estimate for ω is still appropriate. Among 7 non-zero frequency modes, representative are bending vibration at ω=508 cm$^{-1}$ (~0.063 eV) and stretching vibration at ω=816 cm$^{-1}$ (~0.101 eV). At bending vibration, the B atoms coordinates are; B1(1/3+d,0,0), B2(2/3+d,0,0), B3(0,1/3+d,0), B4(0,2/3+d,0), B5(1/3-d,1/3-d,0), B6(1/3,2/3,0), B7(2/3,1/3,0), B8(2/3-d,2/3-d,0), with d standing for displacement out of equilibrium. For stretching vibration mode the coordinates are; B1(1/3-d,0,0), B2(2/3+d,0,0), B3(-d,1/3,0), B4(d,2/3,0), B5(1/3+d,1/3,0), B6(1/3,2/3-d/2,0), B7(2/3,1/3+d/2,0), B8(2/3-d,2/3,0), respectively B1(1/3,d,0), B2(2/3,-d,0), B3(0,1/3+d,0), B4(0,2/3-d,0), B5(1/3,1/3-d,0), B6(1/3-d,2/3-d,0), B7(2/3+d,1/3+d,0), B8(2/3,2/3+d,0).

### 3.2 Band structure calculations of boron tubular polymorph

With helical parameters (*23,0*) and $m_{tr}$=91 the Bα8-SWBNT is generated. Relaxed, i.e. optimized equilibrium lattice parameters are slightly changed comparing to the parent structure; now it is a=b=5.2950 Å. Diameter of the tube is Ø=38.74 Å and the length is ~417 Å that corresponds to experiment [10]. In Fig. 3a we present calculated BS in selected paths (0,0) →(0,1/2) and [($r/m_a$,1/2)→($r/m_a$,0) | (-$r/m_a$,0) →(-$r/m_a$,1/2)] for $r$ = ±1,…,±4 that are counterparts of Γ-M in Bα8 for defined - allowed $k_{Φr}$. As it can be expected for a large diameter SWBNT, for $k_{Φr}$= 0 the BS topology is nearly identical with that in Γ-M path for Bα8 sheet (cf. Fig.1a). Complete set of the BS for full range of allowed $k_{Φr}$, r= 0,±1,…,±11, is presented in Supplementary Data (SD1.1).

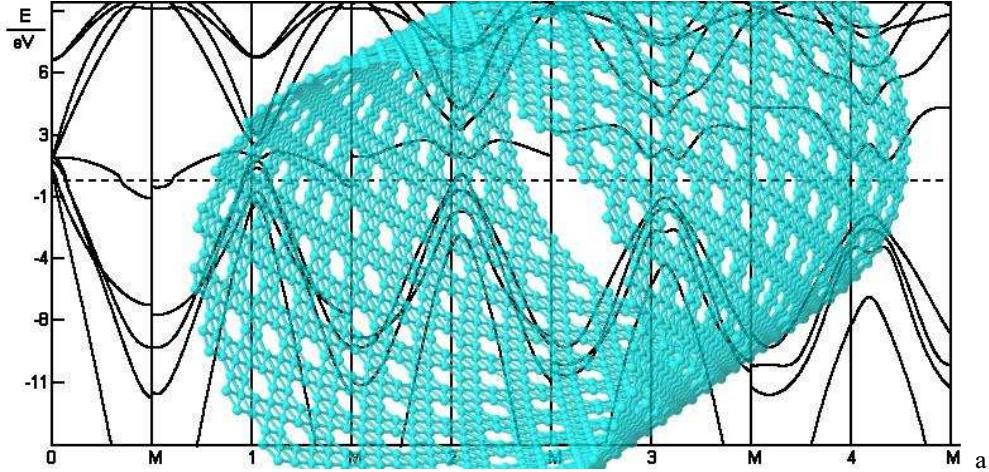

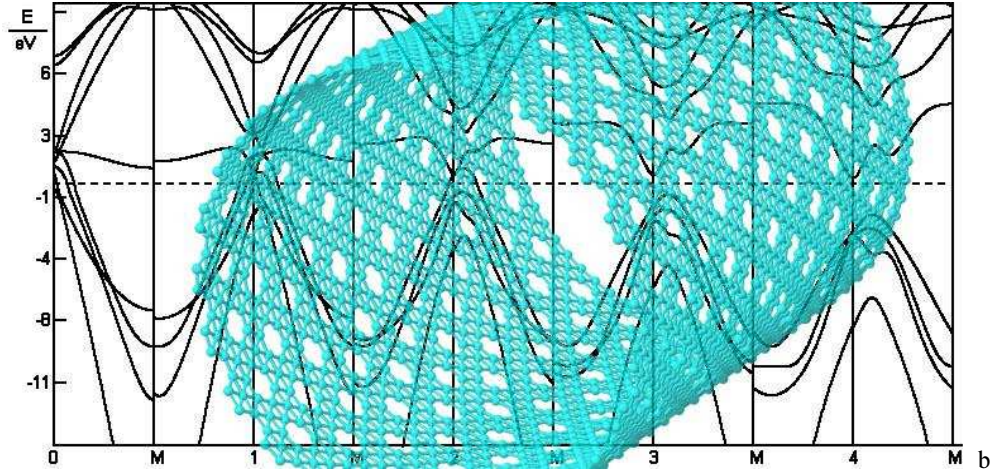

**Figure3. (colour online)** Partial band structure for Bα8-SWBNT in equilibrium (a) and distorted geometry in stretching mode vibration (b). Equivalents of the Γ-M paths for $k_{\phi r}=(r/m_a)$, $r = 0,\pm1,\ldots,\pm4$ are displayed (see the text). The full range BS, $r = 0,\pm1,\ldots,\pm11$ is shown in the SD1.1. Bα8-SWBNT structural motif is in the background.

Chemical potential derived from the BS is $\mu=E_F\approx 0.4$ eV. The ratio $\omega/E_F\approx 0.16$–$0.25$ indicates that Bα8-SWBNT in equilibrium geometry is an adiabatic system.

Influence of electron-vibration coupling on electronic structure is studied for distorted geometries with B-atom displacements in particular vibration mode. Whilst coupling to the bending mode leaves the BS topology unchanged, coupling to the stretching mode implies significant changes at the FL (see Fig. 3b). For a fractional B-displacements of d=0.002, i.e. 0.0106 Å/B from the equilibrium, minimum of the $p_z$ band in M point is shifted above the FL and the top of one of the σ bands in Γ(0) point is shifted below the FL. At a vibration in this mode it represents a periodic fluctuation (cf. Fig. 3a↔b, $r =0, \pm1$) of analytic critical point (ACP) of $p_z$ and the lowest σ bands across FL, which reduces the chemical potential ($\mu\to0$) and induces transition of the system into an anti-adiabatic state. Our results indicate that the effective transition into anti-adiabatic state is present only for Bα8-SWBNTs with diameter larger than 15 Å, i.e. starting with tubular structure of helical vector (9,0) – see SD1.1.1.

Similar results are obtained for Bα8Mg-SWBNT, where Mg atoms are incorporated into tubular structure. The BS for *(23,0)* tube (Ø=39.6 Å, length 426.6 Å) in equilibrium geometry (relaxed structure, a=b=5.407 Å and c=2.8107 Å) for paths as in Fig 3 is shown in Fig. 4a.

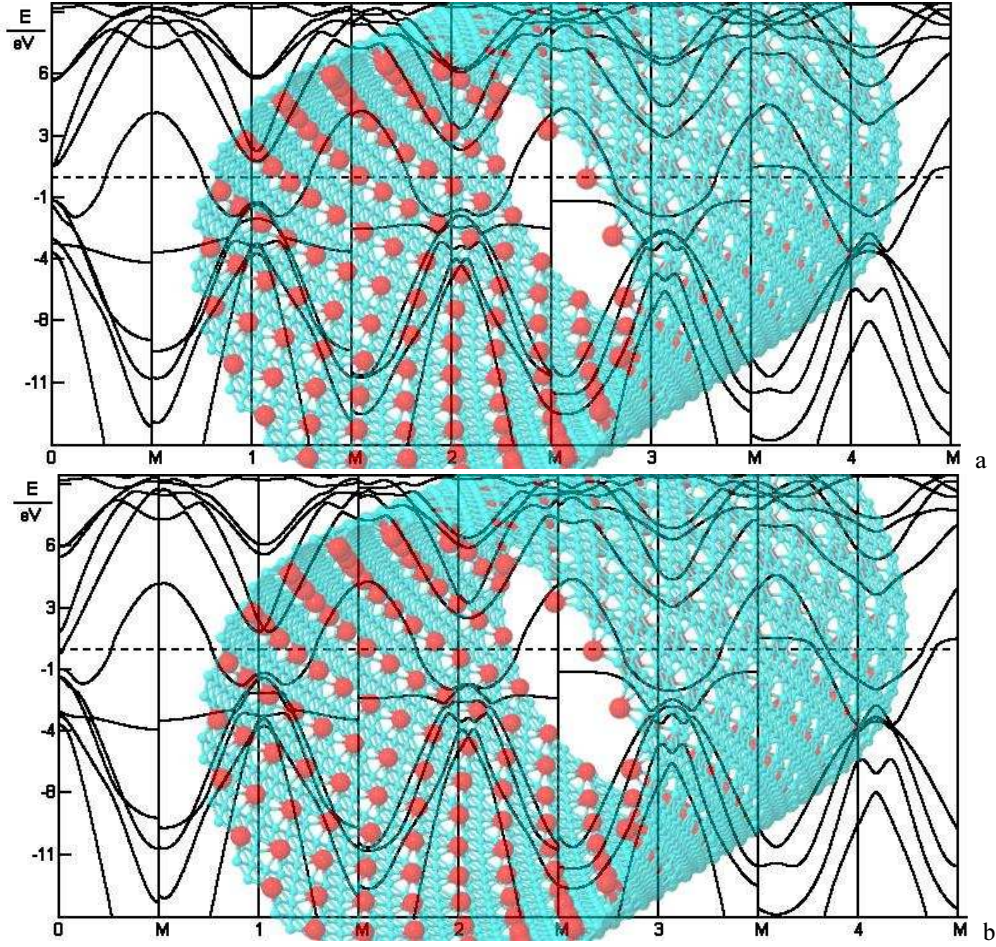

**Figure 4. (colour online)** Partial band structure for Bα8Mg-SWBNT in equilibrium (a) and distorted geometry in stretching mode vibration (b). Equivalents of the Γ-M paths for $k_{\Phi r}=(r/m_a)$, $r = 0,\pm 1,\ldots,\pm 4$ are displayed (see the text). The full range BS, $r = 0,\pm 1,\ldots,\pm 11$ is shown in SD2.1. Bα8Mg-SWBNT structural motif is in the background.

For $k_{\Phi r}=0$, there is again high similarity with the BS topology of Bα8Mg-sheet in Γ-M direction (cf. Fig. 1b). In analogy to Bα8-SWBNT, with chemical potential $\mu=E_F \approx 0.6$ eV, the system is adiabatic ($\omega/E_F \approx 0.1-0.17$) in equilibrium geometry. Coupling to the bending mode leaves the topology of the BS unchanged and, stretching vibration displacement (d=0.002, i.e. 0.0108 Å/B) induces changes of the BS topology at FL – minimum of the conduction band at Γ(0) point is now merged below FL, but maximum of the $p_z$ band remains above the FL in M point for $k_{\Phi r}=0$ (Fig. 4b). Fluctuation (cf. Fig 4a↔b) of the ACP triggers transition of the system into an anti-adiabatic state. It should be noticed that in case when Mg atoms are placed on the outer side of the tubular structure, coupling to the stretching mode does not induce anti-adiabatic state transition – see SD2.2. Replacement of Mg atoms for Al also suppresses transition into anti-adiabatic state and tube in electron-vibration coupling remains metallic – see SD2.3.

Basically the same results, as far as BS fluctuation is concerned, have been obtained for tubular structures with helical vector (*23,23*). The difference is only in appearance of tubular structure. For this type of helical vector, the ends of tube are oblique.

### 3.3 Anti-adiabatic state

An energy increase $\Delta E_d$ is characteristic for nuclear displacements with respect to the minimum on the potential energy surface (crude-adiabatic level). For the aforementioned stretching mode displacements that induced BS fluctuations, $\Delta E_d$ varied up to +18 meV for Bα8-SWBNT and up to +40 meV for Bα8Mg-SWBNT. Going beyond the BOA by inclusion of the nuclear dynamics at EP coupling [25,26] provides corrections to ground state energy ($\Delta E_{na}^0$)-eq.6, to orbital energies ($\Delta \varepsilon_k$)-eq.8a,b and to the electron correlation energy –eq.13. The $\Delta E_{na}^0$ in anti-adiabatic regime ($\omega \geq E_F$) can be negative in dependence on the strength of the non-adiabatic e-p interactions ($u_{C_k,D_{k'}}^s$) and/or the densities ($n_{\varepsilon^0(C_k)}, n_{\varepsilon^0(D_{k'})}$) of interacting states ($\varphi_k, \varphi_{k'}$) mediated by a particular phonon mode (s).

For Bα8-SWBNT, there are 2 fluctuating bands (σ, $p_z$) that interact with 3 other σ bands intersecting the FL whose densities of states (DOS) $n^0(\varepsilon_k^0) = |(\partial k / \partial \varepsilon_k^0)|_{FL}$ are constant and very low at FL, ≈ 0.03-0.05. When the fluctuating bands approach FL for vibration displacements of d≈0.01 Å, DOS increases substantially mainly for the flat $p_z$ band at M point (Fig. 5a). Calculated mean value of $\tilde{u}$ for the interaction of fluctuating bands ($p_z$,σ) with 3 neighbouring σ bands at d=0.01 Å is 1.0 eV. For $n_{pz}≈1.2$ (at ω/2 eV from the FL) and $n_σ≈0.3$, the resulting mean value of $\Delta E_{na}^0$ is ≈ -48 meV that stabilises Bα8-SWBNT by $E_{cond}^0$ ~ -30 meV at distorted geometry in anti-adiabatic ground state. Moreover, anti-adiabatic ground state is geometrically degenerate, with fluxional B-atoms geometry in the stretching displacements (flux-circles with the same radii ~0.01 Å).

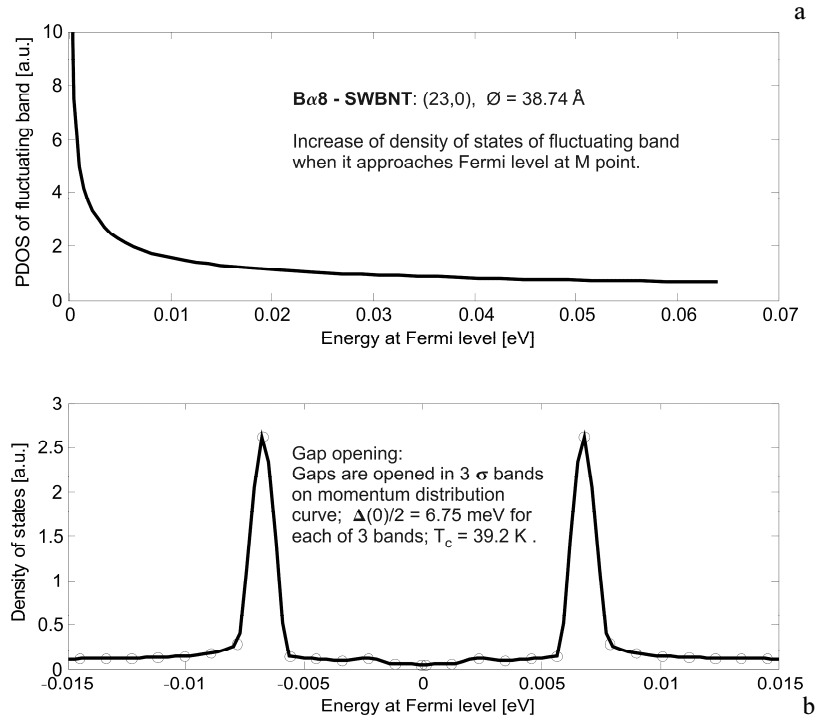

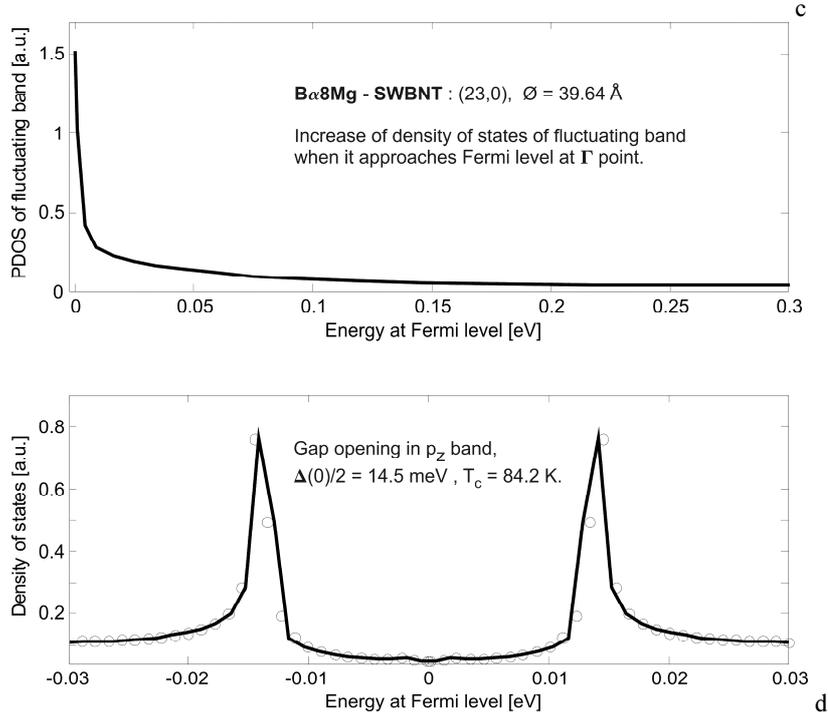

**Figure 5.** Increase of the partial density of states of fluctuating: (a) $p_z$ band at M point in Bα8-SWBNT and (c) conduction band at Γ point in Bα8Mg-SWBNT when the band approaches FL. Gap formation in one-particle spectrum of stabilised anti-adiabatic state due to EP coupling corrections to adiabatic orbital energies is shown in (b) for Bα8-SWBNT and in (d) for Bα8Mg-SWBNT.

For Bα8Mg-SWBNT, the fluctuating conduction band at Γ point increases its DOS at FL (Fig. 5c) and interacts with $p_z$ band that intersects the FL. For mean values of $\tilde{u} \approx 1.37$ eV and $n_c \approx 0.3$, $n_{pz} \approx 0.05$, $\Delta E_{na}^0 \approx -65$ meV stabilises Bα8Mg-SWBNT by $E_{cond}^0 \approx -25$ meV at a distorted geometry in anti-adiabatic ground state with fluxional B-atoms configuration – degenerated ground state.

Orbital energy corrections $\Delta \varepsilon_k$ in anti-adiabatic ground state shift the original adiabatic orbital energies $\varepsilon_k^0$, upwards for unoccupied states at FL, whereas $\varepsilon_k^0$ of the occupied states at FL are shifted downwards – (8a,b). Consequently, the DOS at FL is changed, $n(\varepsilon_k) = \left|1 + \left(\partial(\Delta \varepsilon_k)/\partial \varepsilon_k^0\right)\right|^{-1} \cdot n^0(\varepsilon_k^0)$ and a gap at FL is opened. For Bα8-SWBNT, gaps in three σ bands that intersect the FL in Γ-M direction on adiabatic level are now opened. Three gaps (or a single gap with 2 shoulders) should be distinguished on the momentum distribution curve. On the energy distribution curve a single gap appears, $\Delta(0)/2 \approx 6.7$ meV, as calculated for $n_{pz} \approx 1.2$ at ω/2 eV from the FL and $\tilde{u} = 0.333$ eV/band (Fig. 5b). Single gap of $\Delta(0)/2 \approx 14.5$ meV is opened in $p_z$ band for Bα8Mg-SWBNT (Fig. 5d). An approximate relation (11), $T_c = \Delta(0)/4k_B$, follows for the critical temperature of the adiabatic↔anti-adiabatic state transition from the temperature dependence of an opened gap (7). Mean value of $T_c \approx 39.2$ K results for Bα8-SWBNT. A critical temperature range $T_c \approx$ 70-93.2 K has been calculated for Bα8Mg-SWBNT, depending on $n_c$ (≈0.2-0.4) for an energy range of 0-ω/2 eV at the FL.

## 4. Conclusion

As presented in part 2.2, stabilization of an anti-adiabatic ground state at distorted geometry induces formation of bipolarons – see Eq.21a. As a mobile inter-site electronic polarization (singlet bosons) in real space, due to the fluxional character of B-atoms configuration (geometric degeneracy of the ground state), bipolarons can move on the surface of the tube at $T \leq T_c$ without dissipation in external electric potential. In anti-adiabatic state the thermodynamic properties, including magnetic properties, correspond to thermodynamics of a superconducting state – 2.2d2,d3. Our results show that a transition into the anti-adiabatic state of superconducting character appears for B$\alpha$8-SWBNT and B$\alpha$8Mg-SWBNT with a diameter greater than 15 Å, i.e. for the helical parameter $m_a \geq 9$. In B$\alpha$8Mg-SWBNT, an anti-adiabatic state transition is detected merely for structures with Mg-atoms placed inside the tube. Replacing Mg by Al gives rise to suppressing of an anti-adiabatic state transition and B$\alpha$8Al-SWBNT remains metallic down to 0 K – see SD2.3.

To make the final conclusion on high temperature superconductivity in boron nanotubes requires a more extensive experimental study to be done. Our theoretical results predict that SWBNT should be in this respect very interesting and perspective material.

## Acknowledgements

This work was partially supported by a research grant VEGA No. 1/0005/11 and No. 2/0079/09 of the Ministry of Education of the Slovak Republic. The authors acknowledge Uniqstech a.s. for permission to use the source code of Solid2000 for this research.

## Supplementary Data

The full set BS for allowed values of rotation wave numbers $r = 0,\pm 1,\ldots,\pm 11$ of studied SWBNTs is presented in SD1-2. The BS calculations were performed using cyclic cluster approach at semiempirical INDO level [29] employing the helical symmetry as described in the Section 2.1.

# Supplementary Data
___________________________________________________________________________

# Toward possibility of high-temperature bipolaronic superconductivity in boron tubular polymorph: Theoretical aspects of transition into anti-adiabatic state

Pavol Baňacký[1*], Jozef Noga[2,3] and Vojtech Szöcs[1]

[1]Chemical Physics division, Institute of Chemistry, Faculty of Natural Sciences, Comenius University, Mlynska dolina CH2, 84215 Bratislava, Slovakia

[2]Department of Inorganic Chemistry, Faculty of Natural Sciences, Comenius University, Mlynska dolina CH2, 84215 Bratislava, Slovakia

[3]Institute of Inorganic Chemistry, Slovak Academy of Sciences, 84536 Bratislava, Slovakia

- **SD1.** Band structures of Bα8 tubular structures with respect to adiabatic ↔ anti-adiabatic state transition at e-p coupling

  **SD1.1**. Band structure of Bα8 tube: helical vector (23,0), diameter 38.74 Å, length 417,02 Å, for the full set of rotation "wave vector" $k_\Phi$

  **SD1.1.1.** Band structure of Bα8 tubes with small diameter:

  **a/** Bα8 tube, helical vector (9,0), diameter 15.19 Å, length 417,02 Å, for the full set of rotation "wave vector" $k_\Phi$

  **b/** Bα8 tube, helical vector (7,0), diameter 11.79 Å, length 417,02 Å, for the full set of rotation "wave vector" $k_\Phi$

  **c/** Bα8 tube, helical vector (5,0), diameter 8.42 Å, length 417,02 Å, for the full set of rotation "wave vector" $k_\Phi$

- **SD2.1.** Band structure of Bα8Mg tube (Mg inside) with adiabatic ↔ antiadiabatic state transition at e-p coupling: helical vector (23,0), diameter 39.64 Å, length 426.65 Å, for full set of rotation wave vector $k_\Phi$

- **SD2.2.** Band structure of Bα8Mg tube (Mg outside) which remains metallic at e-p coupling: helical vector (23,0), diameter 39.64 Å, length 426.65 Å, for the full set of rotation wave vector $k_\Phi$

- **SD2.3.** Band structure of Bα8Al tube (Al inside) which remains metallic at e-p coupling: helical vector (23,0), diameter 39.64 Å, length 426.65 Å, for different values of rotation wave vector $k_\Phi$

---

[*] corresponding author, banacky@fns.uniba.sk

# SD1. Band structures of Bα8 tubular structures with respect to adiabatic ↔ anti-adiabatic state transition in EP coupling

## SD1.1. Band structure of Bα8 tube: helical vector (23,0), diameter 38.74 Å, length 417,02 Å, for full set of rotation "wave vector" $k_\Phi$

Rotation "wave vector": $k_\Phi=(r/m_a)$, $m_a=23$; $r=0,\pm1,\pm2,\ldots,\pm(m_a-2)/2,\pm(m_a-1)/2$; M is high symmetry point of hexagonal primitive lattice, M:[0,1/2,0]

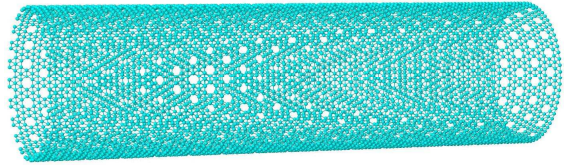

**At equilibrium nuclear configuration**
r = (0 - ± 4)

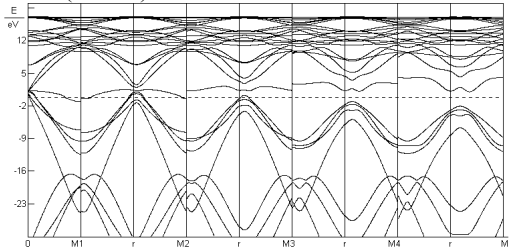

**At displacements in stretching normal mode ω=816 cm$^{-1}$**
r = (0 - ± 4)

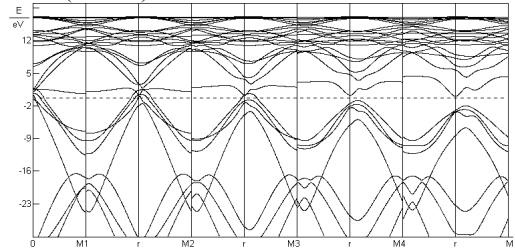

**At equilibrium nuclear configuration**
r = (± 5 - ± 8)

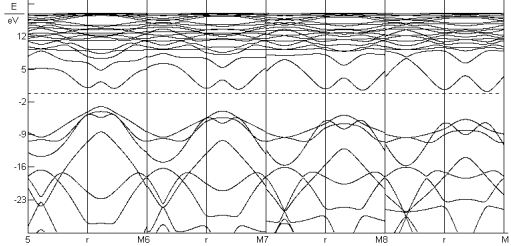

**At displacements in stretching normal mode ω=816 cm$^{-1}$**
r = (± 5 - ± 8)

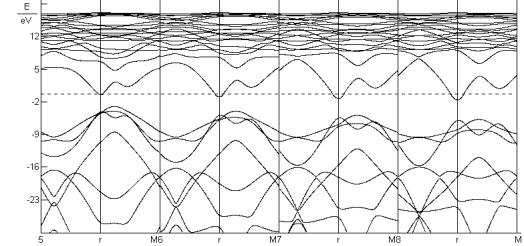

**At equilibrium nuclear configuration**
r = (± 9 - ± 11)

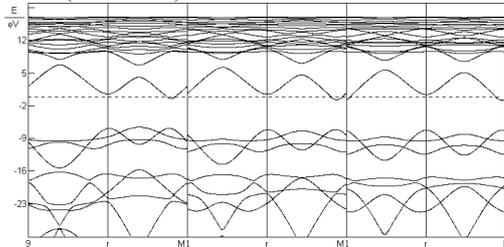

**At displacements in stretching normal mode ω=816 cm$^{-1}$**
r = (± 9 - ± 11)

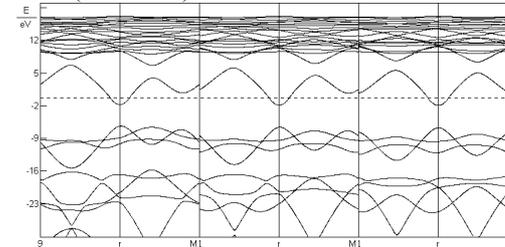

Coupling to vibration mode induces fluctuation of $p_z$ band at M-point and σ band at Γ point. System is stabilized in anti-adiabatic state by $E^0_{cond} \approx$ -30 meV ($\Delta E^0_d \approx$ +18 meV  $\Delta E^0_{na} \approx$ -48 meV).

### SD1.1.1. Band structure of Bα8 tubes with small diameter: Comparison of B-tubes with diameter > 15 Å which undergoes adiabatic↔antiadiabatic state transition at e-p coupling, with B-tubes with diameter < 15 Å which remain metallic down to 0K.

**a/ Bα8 tube, helical vector (9,0), diameter 15.19 Å, length 417,02 Å, for full set of rotation "wave vector" $k_\Phi$**

Rotation "wave vector": $k_\Phi = (r/m_a)$, $m_a = 23$; $r = ,0,\pm1,\pm2,\ldots,\pm(m_a-2)/2,\pm(m_a-1)/2$; M is high symmetry point of hexagonal primitive lattice, M:[0,1/2,0]

| At equilibrium nuclear configuration | At displacements in stretching normal mode ω=816 cm$^{-1}$ |
|---|---|
| r = (0,±1,±2,±3,±4) | r = (0,±1,±2,±3,±4) |

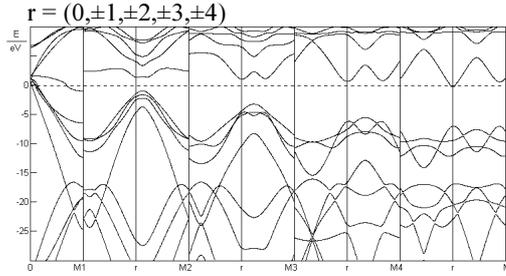 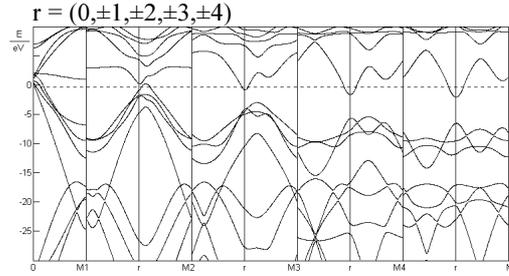

**b/ Bα8 tube, helical vector (7,0), diameter 11.79 Å, length 417,02 Å, for full set of rotation "wave vector" $k_\Phi$**

| At equilibrium nuclear configuration | At displacements in stretching normal mode ω=816 cm$^{-1}$ |
|---|---|
| r = (0,±1,±2,±3) | r = (0,±1,±2,±3) |

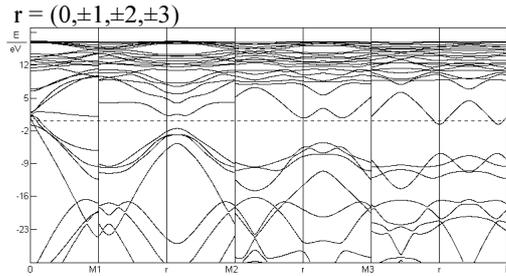 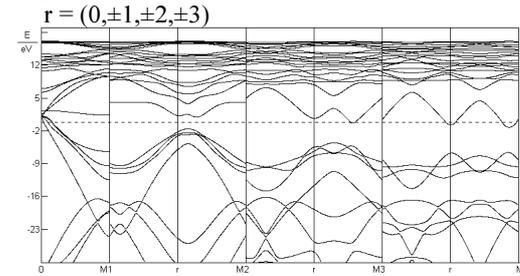

As it can be seen for (9,0) tube-/a, there is fluctuation of $p_z$ band (at M-point) and σ band (at Γ-point) across FL in EP coupling. Electronic structure of the tube exhibits similar effects like (23,0) tube. Calculated $\Delta E_{na}$ is great enough ($|\Delta E_{na}| > \Delta E_d$) to stabilize system in anti-adiabatic state, mainly due to contribution of $p_z$ band. In case of (7,0) tube-/b, the only σ band (at Γ-point) fluctuates across FL. Calculated $\Delta E_d$ at crossing is +34 meV. Contribution of correction to ground state energy $\Delta E_{na}$ due to nonadiabatic e-p coupling is ~ - 5 meV (calculated for $\tilde{u} \approx 0.37$ eV, $n_\sigma \approx 0.3$ for fluctuating σ band and $n_\sigma \approx 0.03$/band for 3 other bands at FL). This value is not great enough ($|\Delta E_{na}| < \Delta E_d$) to stabilize the system in anti-adiabatic state and (7,0) tube remains metallic.

**c/ Bα8 tube, helical vector (5,0), diameter 8.42 Å, length 417,02 Å, for full set of rotation "wave vector" $k_\Phi$**

| At equilibrium nuclear configuration | At displacements in stretching normal mode ω=816 cm$^{-1}$ |
|---|---|
| r = (0,±1,±2,±3) | r = (0,±1,±2,±3) |

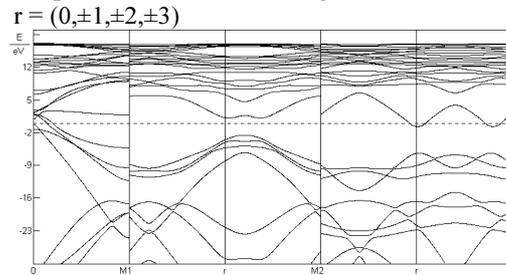 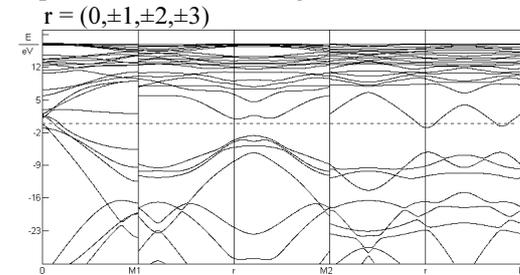

In this case, coupling to vibration mode does not induce fluctuation of bands across FL. Tube is metallic.

**SD2.1 Band structure of Bα8Mg tube (Mg inside) with adiabatic ↔ antiadiabatic state transition in EP coupling: helical vector (23,0), diameter 39.64 Å, length 426.65 Å, for full set of rotation wave vector $k_\Phi$**

Rotation wave vector: $k_\Phi = (r/m_a)$
$m_a = 23$; $r = 0, \pm 1, \pm 2, \ldots, \pm(m_a-2)/2, \pm(m_a-1)/2$; M is high symmetry point of hexagonal primitive lattice, M:[0,1/2,0]

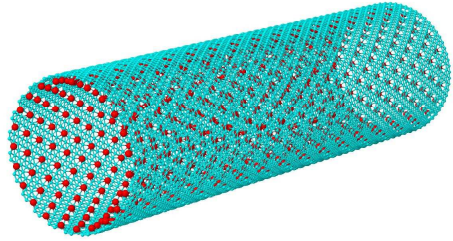

**At equilibrium nuclear configuration**
r = (0 - ± 4)

**At displacements in stretching normal mode ω=816 cm⁻¹**
r = (0 - ± 4)

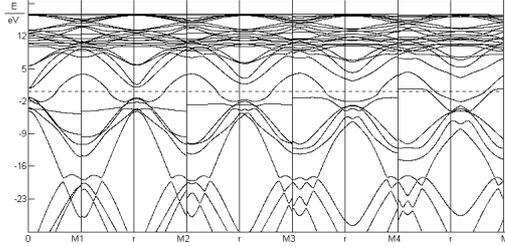 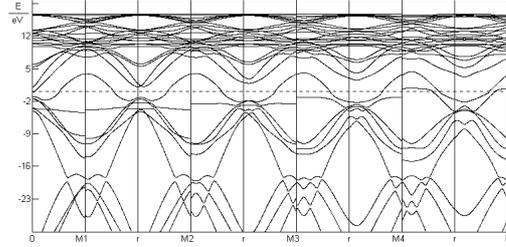

**At equilibrium nuclear configuration**
r = (± 5 - ± 8)

**At displacements in stretching normal mode ω=816 cm⁻¹**
r = (± 5 - ± 8)

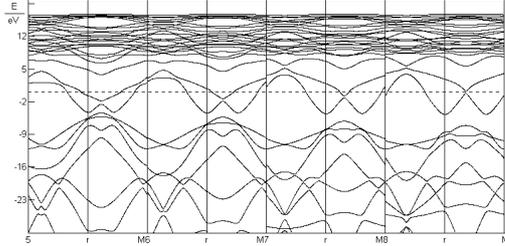 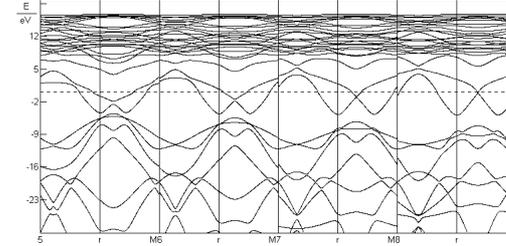

**At equilibrium nuclear configuration**
r = (± 9 - ± 11)

**At displacements in stretching normal mode ω=816 cm⁻¹**
r = (± 9 - ± 11)

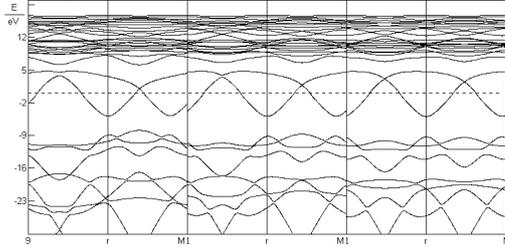 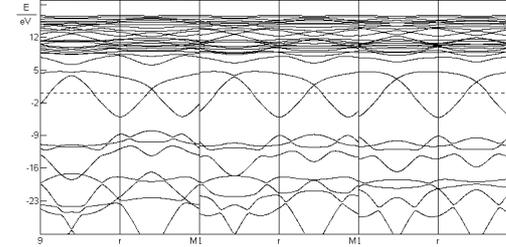

In this case, coupling to vibration mode induces fluctuation of σ band at Γ point. System is stabilized in anti-adiabatic state by $E^0_{cond} \approx$ -25 meV ( $\Delta E^0_d \approx$ +40 meV  $\Delta E^0_{na} \approx$ -65 meV).

**SD2.2 Band structure of Bα8Mg tube (Mg outside) which remains metallic in EP coupling: helical vector (23,0), diameter 39.64 Å, length 426.65 Å, for full set of rotation wave vector $k_\Phi$**

Rotation wave vector: $k_\Phi = (r/m_a)$
$m_a = 23$; $r = ,0, \pm 1, \pm 2, \ldots, \pm(m_a-2)/2, \pm(m_a-1)/2$; M is high symmetry point of hexagonal primitive lattice, M:[0,1/2,0]

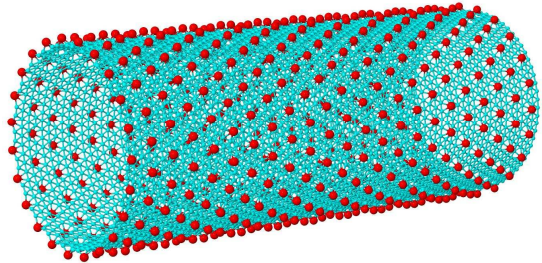

**At equilibrium nuclear configuration**  **At displacements in stretching normal mode ω=816 cm⁻¹**
r = (0 - ± 4)                              r = (0 - ± 4)

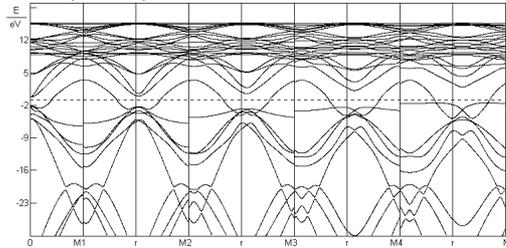 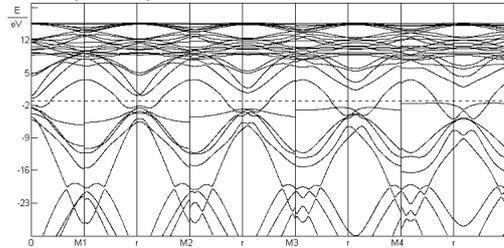

**At equilibrium nuclear configuration**  **At displacements in stretching normal mode ω=816 cm⁻¹**
r = (± 5 - ± 8)                            r = (± 5 - ± 8)

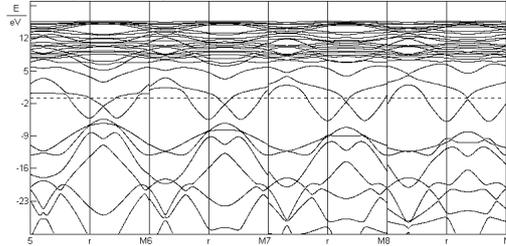 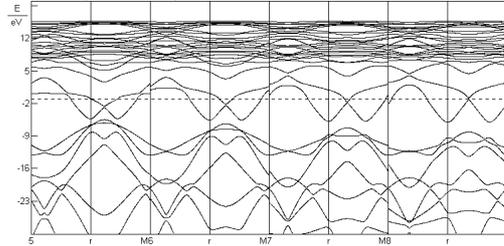

**At equilibrium nuclear configuration**  **At displacements in stretching normal mode ω=816 cm⁻¹**
r = (± 9 - ± 11)                           r = (± 9 - ± 11)

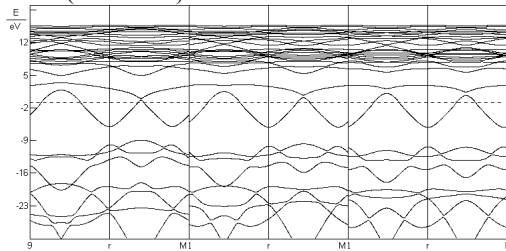 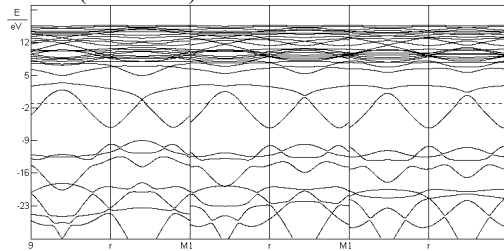

In this case, coupling to vibration mode does not induce fluctuation of bands across FL. Tube is metallic.

**SD2.3 Band structure of Bα8Al tube (Al inside) which remains metallic at e-p coupling: helical vector (23,0), diameter 39.64 Å, length 426.65 Å, for different values of rotation wave vector $k_\Phi$**

Rotation wave vector: $k_\Phi = (r/m_a)$, $m_a = 23$; $r = 0, \pm 1, \pm 2, \ldots, \pm(m_a-2)/2, \pm(m_a-1)/2$; M is high symmetry point of hexagonal primitive lattice, M:[0,1/2,0]

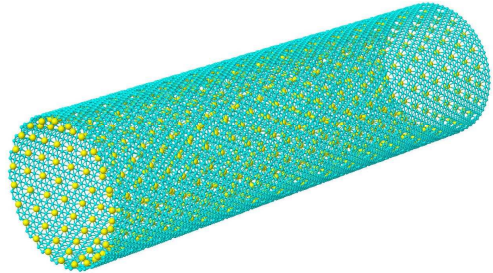

| At equilibrium nuclear configuration | At displacements in stretching normal mode ω=816 cm⁻¹ |
| --- | --- |
| r = (0 - ± 4) | r = (0 - ± 4) |

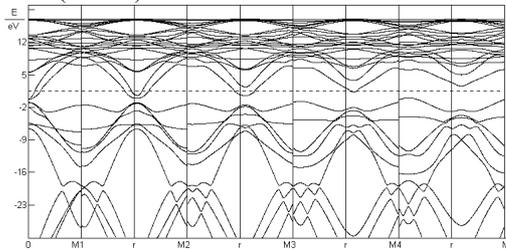 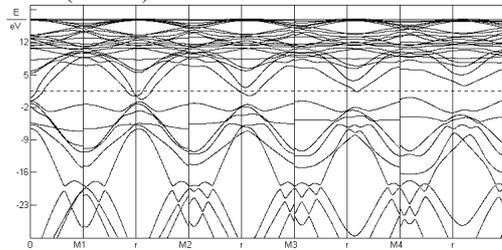

In this case, coupling to vibration mode does not induce fluctuation of bands across FL. Tube is metallic.